\documentclass[preprint,12pt]{elsarticle}

\usepackage{amssymb}
\usepackage{amsmath}
\usepackage{graphicx}
\usepackage{booktabs}
\usepackage{float}
\usepackage{hyperref}
\usepackage{natbib}

\journal{Journal of Economic Interaction and Coordination}

\begin{document}

\begin{frontmatter}

\title{Order Splitting and Liquidity Replenishment Are Jointly
Necessary for the Square-Root Law of Market Impact:\\
A Counterfactual Dissection}

\author[ins,physics]{Yang Zhou\corref{cor}}
\cortext[cor]{Corresponding author}
\ead{zhouyang@westlake.edu.cn}

\author[tobacco]{Jianwen Chen}
\author[swufe]{Ruipeng Wei}

\address[ins]{Institute of Natural Sciences, Westlake Institute for Advanced Study,
Hangzhou 310024, China}
\address[physics]{Department of Physics, School of Science and Research Center
for Industries of the Future, Westlake University, Hangzhou 310030, China}
\address[tobacco]{Greatwall Cigar Factory of China Tobacco Sichuan Industrial Co., Ltd., Sichuan, China}
\address[swufe]{Southwestern University of Finance and Economics, Chengdu, Sichuan, China}

\begin{abstract}
Three quantitative predictions have been advanced for the square-root law
(SRL) of market impact, $I/\sigma_D = c\,(Q/V_D)^{\delta}$ with
$\delta\approx 0.5$: GGPS ($\delta=\beta-1$), FGLW
($\delta=\alpha-1$), and LOB walking ($\delta=1/(1+\gamma)$). Using a
minimal limit-order-book model populated by heterogeneous interacting
agents and calibrated against the Tokyo Stock Exchange benchmark
($\langle\delta\rangle = 0.489$~\citep{satoStrictUniversalitySquareRoot2025}),
we test all three on identical simulated data and find that none matches
the per-stock measured $\delta$: GGPS and FGLW over-predict by factors of
two and four respectively, while LOB walking under-predicts. The model
reproduces $\langle\delta\rangle = 0.539\pm 0.048$ across 2000
independently parameterised stocks. To identify which mechanisms are
causally responsible, we perform counterfactual ablation by selectively
suppressing each component. Removing order splitting collapses $\delta$
from $0.549$ to $0.324$; removing liquidity replenishment by market
makers drops it to $0.386$; perturbations that leave both intact
(momentum trading, price limits, splitting rule, background liquidity)
move $\delta$ by less than $10\%$. Order splitting and liquidity
replenishment are thus jointly identified as the necessary mechanisms for
the SRL within this model, with the simulated SRL depending on neither
the metaorder size tail nor the visible book shape in isolation.
\end{abstract}


\end{frontmatter}

%% =====================================================================
\section{Introduction}
%% =====================================================================

When a large institutional order (metaorder) of size $Q$ is executed, the
resulting price impact $I$ obeys a power-law scaling:
\begin{equation}
  \frac{I}{\sigma_D} = c \left(\frac{Q}{V_D}\right)^{\delta},
  \label{eq:srl}
\end{equation}
with $\delta \approx 0.5$ across equity markets~\citep{almgrenDirectEstimationEquity2005, moroMarketImpactTrading2009,
tothAnomalousPriceImpact2011, bouchaudTradesQuotesPrices2018, satoStrictUniversalitySquareRoot2025}. Here $V_D$ and $\sigma_D$
denote daily volume and daily volatility respectively. This ``square-root
law'' (SRL) has become an established regularity of quantitative market
microstructure~\citep{bouchaudFluctuationsResponseFinancial2004, contEmpiricalPropertiesAsset2001}, yet the mechanisms that
produce it remain unsettled.

Four theoretical routes have been proposed. The \emph{LOB walking}
argument~\citep{weberOrderBookApproach2005} notes that a market order
crossing a book with cumulative profile $V(q)\sim q^{1+\gamma}$ produces
$\delta = 1/(1+\gamma)$; a parabolic book ($\gamma=1$) gives $\delta=0.5$.
This prediction, however, depends on the \emph{visible} book shape and
ignores latent orders that may appear during execution; Weber \& Rosenow
themselves showed that dynamic feedback reduces actual impact well below
the static-book prediction. The \emph{GGPS} framework of Gabaix et al.~\citep{gabaixInstitutionalInvestorsStock2006}
links $\delta$ to the tail exponent $\beta$ of the metaorder size
distribution via $\delta = \beta-1$, assuming optimal execution --- but
the distribution alone need not determine impact if execution dynamics
mediate the size--displacement relationship.
The \emph{FGLW} theory of Farmer et al.~\citep{farmerHowEfficiencyShapes2013} replaces $\beta$ with the child-order
count tail exponent $\alpha$, predicting $\delta = \alpha-1$; it shares the
same limitation. The \emph{latent liquidity} theory~\citep{mastromatteoAgentbasedModelsLatent2014, donierFullyConsistentMinimal2015} takes a
different route: as execution moves the price, new hidden orders are
continuously revealed, yielding $\delta\approx 0.5$ irrespective of the
visible LOB shape. This framework is elegant but difficult to test
directly, since latent liquidity is by definition unobservable.

These theories make distinct, testable predictions for $\delta$, and a
natural null hypothesis is that one of them captures the relevant
mechanism: that the SRL arises either from the metaorder size distribution
(GGPS), from the child-order count distribution (FGLW), or from the shape
of the visible limit order book (LOB walking). Existing agent-based
studies have reproduced the SRL~\citep{tothAnomalousPriceImpact2011, donierFullyConsistentMinimal2015,
mastromatteoAgentbasedModelsLatent2014}, but none has
systematically disentangled the causal role of each mechanism through
controlled ablation, nor ruled out the competing predictions on the same
simulated data.

The present paper tests this null hypothesis directly. We build a minimal
limit-order-book ABM populated by heterogeneous interacting agents ---
simple enough that each mechanism can be turned on and off --- and use it
in two complementary ways. First, counterfactual ablation: by selectively
suppressing order splitting, removing market makers, imposing price
limits, or varying liquidity, we measure how $\delta$ responds and isolate
the mechanisms whose removal breaks the SRL. Second, a per-stock
computational stress test of the three quantitative predictions (GGPS,
FGLW, LOB walking): for each of 2000 independently simulated stocks we
measure the relevant exponents ($\beta$, $\alpha$, $\gamma$) and compare
predicted versus measured $\delta$. The two lines of evidence converge:
all three predictions are rejected on identical data, and only the
ablation of splitting or of liquidity replenishment substantially
destroys $\delta\approx 1/2$.

%% =====================================================================
\section{Model}
\label{sec:model}
%% =====================================================================

\subsection{Market structure}

The simulated market is a continuous double-auction limit order book.
Both market and limit orders are supported; the tick size is set to
1 basis point of the initial price, mimicking liquid equity markets.

\subsection{Agents}

Four agent types populate the market, each representing a class of real
market participants.

\textbf{Institutional agents} ($N_{\text{inst}}=20$) are the sole source of metaorders.
Each institution draws a total order size $Q$ from a Pareto distribution
with shape $\xi\sim \mathrm{Uniform}[1.5,3.5]$ and splits $Q$ into
$N_c$ child orders that arrive as a Poisson process ($\lambda=3$ ticks).
The default splitting rule is Dirichlet. To ensure the $Q/V_D$ axis spans
the empirically relevant range ($0.01\%$--$10\%$), the minimum order size
$q_{\min}$ is log-spaced across agents from $1$ to $2000$.

\textbf{High-frequency trading (HFT) / market-maker agents} ($N_{\text{HFT}}\in[3,10]$ per stock)
post two-sided limit orders at $\ell=50$ price levels with adaptive spread.
After each fill they replenish with probability $p=0.9$; this provides the
continuous liquidity that lets the LOB recover between child orders.

\textbf{Retail agents} ($N_{\text{retail}}\in[100,400]$ per stock) submit
small, randomly directed market orders of size $[1,10]$. They play the same
role as noise traders in standard microstructure models, providing
background turnover.

\textbf{News agents} ($N=3$) occasionally inject large market orders. Their
purpose is to generate exogenous volatility spikes, without which the
simulated return distribution would lack the heavy tails seen in real
markets.

\subsection{Heterogeneity across stocks}

We simulate $N_s=2000$ stocks with independently drawn parameters
(20 stocks for the counterfactual experiments; see
Section~\ref{sec:counterfactual}). Each stock receives a random initial
price (log-uniform from $[1000,50000]$), a random number of retail traders
($\mathrm{Uniform}[100,400]$), a random number of HFT agents
($\mathrm{Uniform}[3,10]$), and a random Pareto shape $\xi$
($\mathrm{Uniform}[1.5,3.5]$). The tick size scales with the initial price.
This cross-stock variation prevents any single parameter choice from
dominating the results.

\subsection{Simulation protocol}

Each stock runs for $T=10^6$ steps, preceded by a $5\times 10^4$-step
warmup that lets the LOB self-organise. At every step one agent is chosen
uniformly at random to act. Stocks are independent (separate random seeds).
We checked convergence by comparing $\delta$ estimated on the first and
second halves of the baseline trajectory: the two estimates agree to within
$\pm 0.01$ for all 2000 stocks, confirming that $10^6$ steps is sufficient
for the impact exponent to stabilise.
The 2000-stock baseline simulation required approximately 3 hours of
wall-clock time on a 40-core Xeon E5-2698 v4 server (each stock is
embarrassingly parallel).

\subsection{Metaorder identification}

Because the model records every trade, we reconstruct metaorders \emph{ex
post} by grouping consecutive same-agent, same-direction trades. Two trades
from the same agent are assigned to different metaorders if they are
separated by more than $\Delta t = 10$ steps or if the direction flips.
We chose $\Delta t = 10$ because the median inter-child interval in the
baseline is 3 ticks; setting the threshold at roughly three medians
captures the bulk of each metaorder without over-merging distinct orders.

%% =====================================================================
\section{Analysis methodology}
%% =====================================================================

\subsection{Normalization}

We normalise each metaorder by the statistics of the day on which it
executes, following~\citep{satoStrictUniversalitySquareRoot2025}:
\begin{equation}
  Q_{\text{norm}} = \frac{Q}{V_D^{(d)}}, \qquad
  I_{\text{norm}} = \frac{\text{sign} \times I}{\sigma_D^{(d)}},
\end{equation}
where $V_D^{(d)}$ is the day-$d$ total traded volume and
$\sigma_D^{(d)}$ the day-$d$ price range (high$-$low).
Normalising by true per-day quantities rather than aggregate averages
matters because dividing by a constant $V_D/N_{\text{days}}$ only shifts
the log-log plot without changing $\delta$.

\subsection{Fitting procedure}

Logarithmic binning (20 bins, $\geq 30$ points per bin) converts the
scatter of normalised $(Q_{\text{norm}}, I_{\text{norm}})$ pairs into
bin averages, to which we fit $I = c\,Q^{\delta}$ by relative
least squares:
\begin{equation}
  \min_{c, \delta} \sum_i \left(\frac{y_i - c x_i^{\delta}}{y_i}\right)^2.
\end{equation}
Metaorders with $Q/V_D < 0.01\%$ are dropped: at such tiny sizes the
measured impact is dominated by the bid--ask spread rather than by the
shape of the LOB.

%% =====================================================================
\section{Results}
%% =====================================================================

\subsection{Stylized facts}

\begin{figure}[H]
  \centering
  \includegraphics[width=\textwidth]{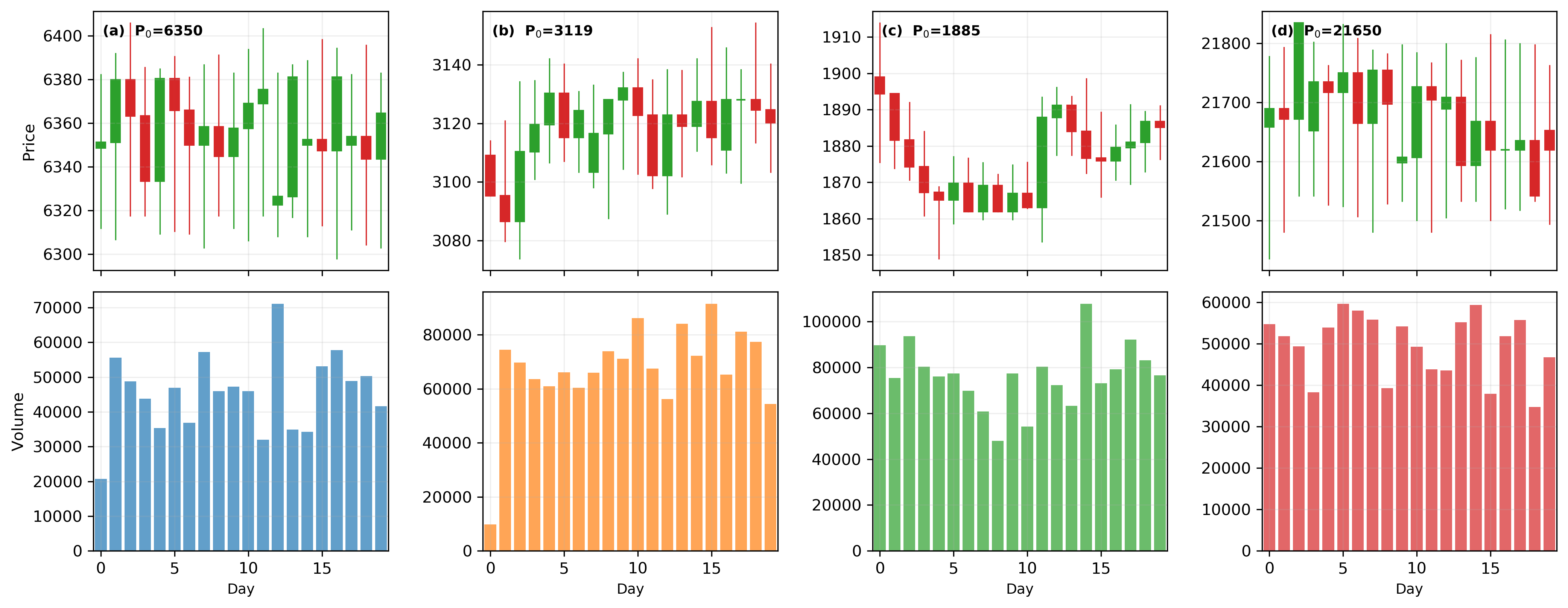}
  \caption{Daily candlestick charts (top) and volume bars (bottom) for
    four stocks. Horizontal axis: trading day; vertical axis: price
    (normalised to $P_0$) and volume (in lots). The four stocks span
    initial prices from $P_0 = 1885$ to $P_0 = 21651$ and exhibit
    different volatility regimes.}
  \label{fig:price_volume}
\end{figure}

Before studying impact, we check that the simulated market reproduces the
basic statistical signatures of real equity markets.
Figure~\ref{fig:price_volume} shows daily candlestick charts and volume for
four stocks with different parameters. Prices behave as random walks with
stock-dependent volatility, as expected.

\begin{figure}[H]
  \centering
  \includegraphics[width=\textwidth]{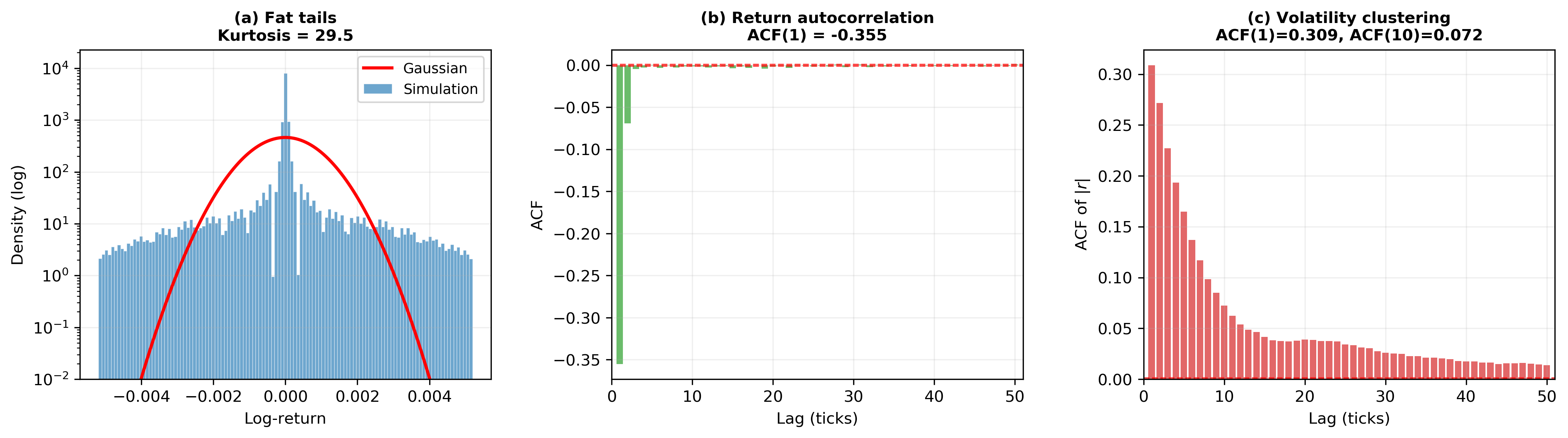}
  \caption{Tick-level return diagnostics for Stock 2 ($P_0 = 1885$).
    (a)~Semi-log density of log-returns (blue circles) vs.\ Gaussian
    with matching $\sigma$ (red line). The heavier tails give excess
    kurtosis $\kappa = 29.5$.
    (b)~ACF of raw returns, lags 1--50. The lag-1 value $-0.355$
    reflects bid--ask bounce~\citep{rollSimpleImplicitMeasure1984}; dashed lines: $\pm 2/\sqrt{N}$.
    (c)~ACF of $|r|$, lags 1--50. The slow decay from $0.309$ at
    lag~1 is the volatility-clustering signature.}
  \label{fig:stylized}
\end{figure}

At tick level (Figure~\ref{fig:stylized}), the simulated returns are
fat-tailed (excess kurtosis $\kappa = 29.5$), negatively autocorrelated at
lag 1 (autocorrelation function ACF(1)$=-0.355$, the classic bid--ask bounce
signature~\citep{rollSimpleImplicitMeasure1984}), and exhibit slowly decaying absolute-return
autocorrelation (ACF(1)$=0.309$), i.e.\ volatility clustering.

\subsection{Square-root law}

\begin{figure}[H]
  \centering
  \includegraphics[width=\textwidth]{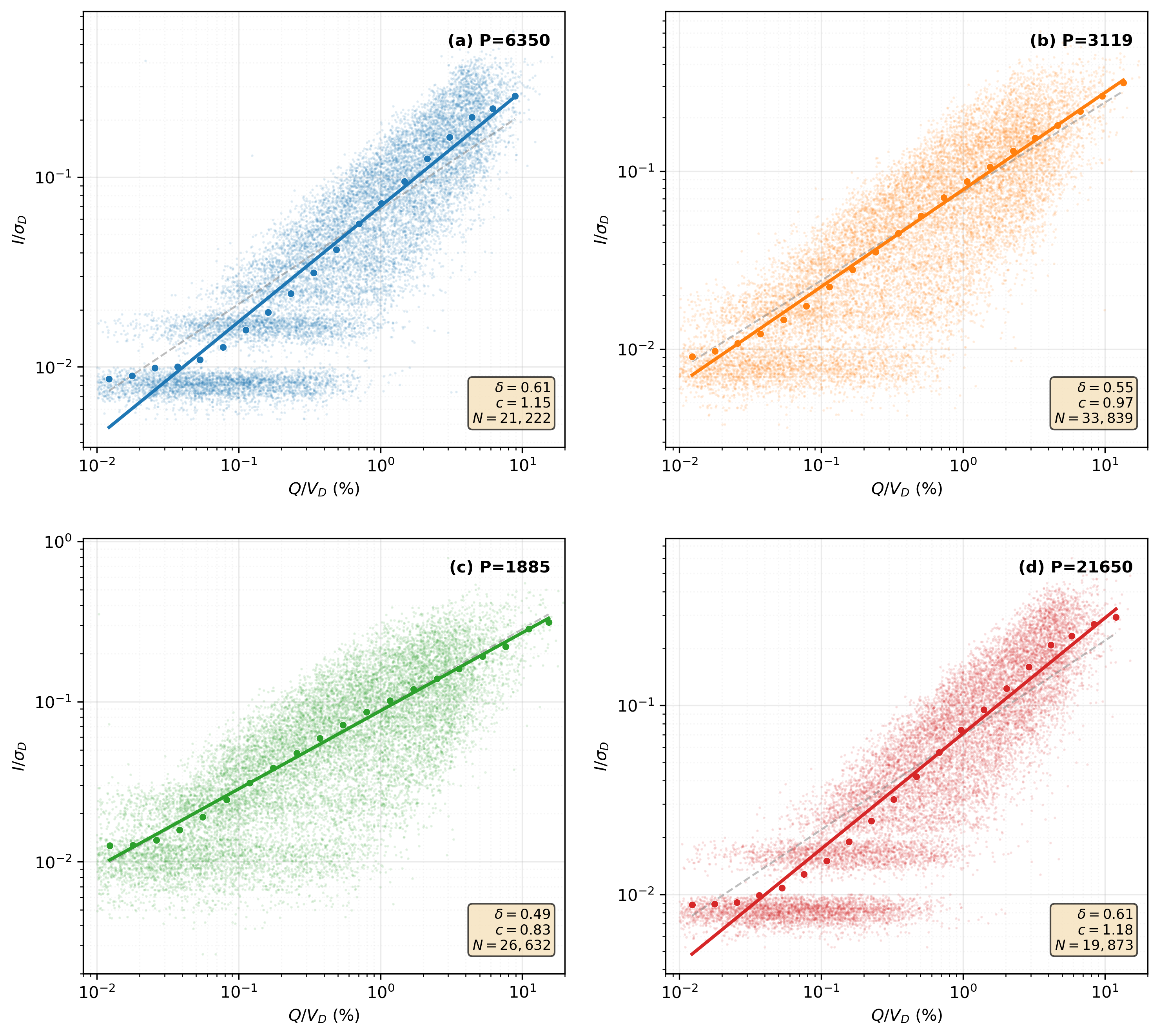}
  \caption{Per-stock impact curves (log-log scale).
    Horizontal axis: $Q/V_D$; vertical axis: $I/\sigma_D$.
    Grey dots: individual metaorders (subsampled). Blue circles:
    log-binned averages (20 bins, $\geq 30$ points each). Red solid:
    power-law fit $I = c\,(Q/V_D)^{\delta}$. Black dashed: $\delta=1/2$
    reference. Each panel reports the fitted $\delta$, $c$, and the
    number of institutional metaorders $N_{\text{meta}}$.}
  \label{fig:per_stock}
\end{figure}

Figure~\ref{fig:per_stock} plots the impact curve for four individual
stocks. Each stock yields a well-defined power law with
$\delta\in[0.49, 0.61]$; all bracket the Tokyo Stock Exchange (TSE) benchmark
$\langle\delta\rangle = 0.489$~\citep{satoStrictUniversalitySquareRoot2025}.

Pooling all stocks gives $\delta = 0.549$, $c = 0.982$, to be compared
with $\langle\delta\rangle = 0.489$, $\langle c\rangle = 0.842$
from~\citep{satoStrictUniversalitySquareRoot2025}.

\begin{figure}[H]
  \centering
  \includegraphics[width=\textwidth]{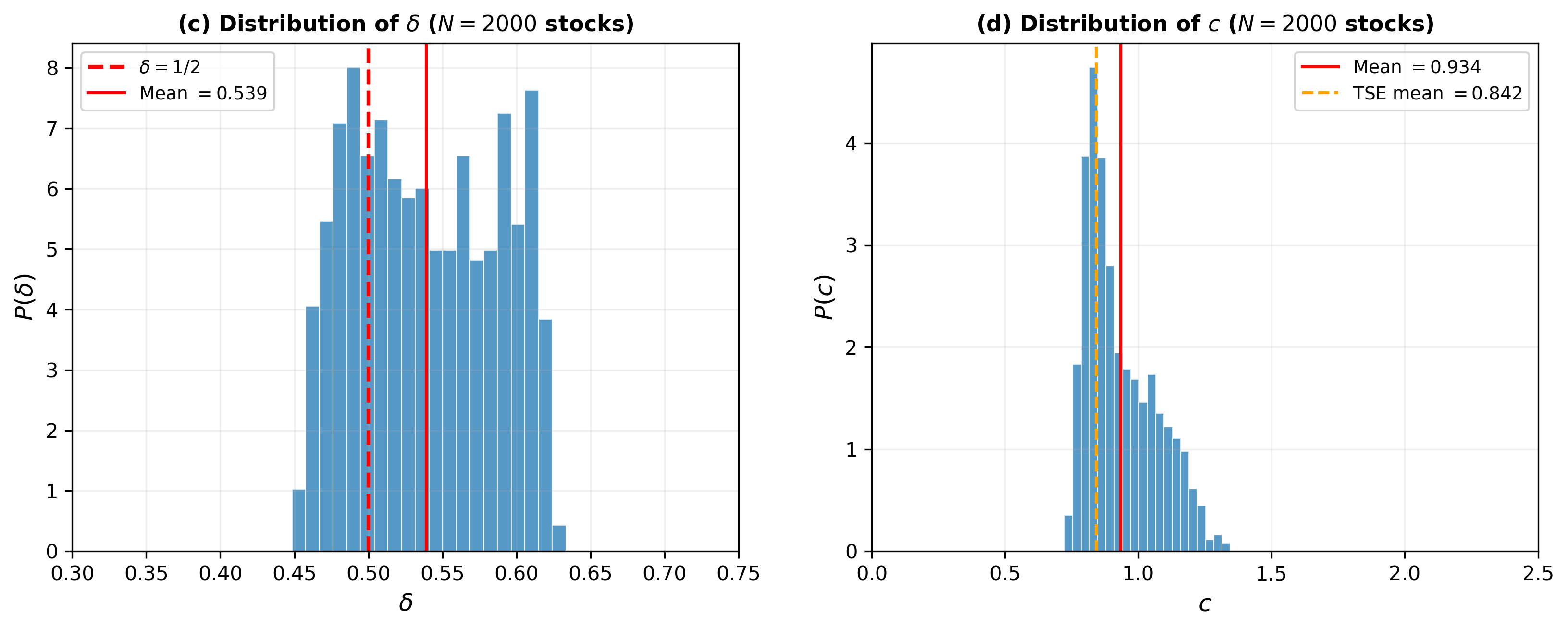}
  \caption{Per-stock fitted parameters from 2000 independent simulations.
    (a)~Histogram of $\delta$ (bin width $= 0.02$). Red solid line:
    cross-stock mean $= 0.539$. Red dashed: $\delta = 1/2$.
    The distribution is roughly symmetric around $0.54$.
    (b)~Histogram of $c$ (bin width $= 0.05$). Red solid: mean $= 0.934$.
    Orange dashed: TSE empirical mean $= 0.842$~\citep{satoStrictUniversalitySquareRoot2025}.}
  \label{fig:delta_distribution}
\end{figure}

To check robustness, we repeat the analysis on 2000 stocks with
independently drawn parameters. The per-stock $\delta$ distribution
(Figure~\ref{fig:delta_distribution}) has mean $0.539\pm 0.048$ and range
$[0.449, 0.633]$, while $\langle c\rangle = 0.934\pm 0.130$.
The TSE distribution spans a wider range ($[0.2, 0.7]$ across 2000
stocks) because real markets vary in trading rules, investor composition,
and regulatory regime --- sources of variation absent from our minimal
setup. Both distributions nonetheless centre on $\delta \approx 1/2$.

\begin{figure}[H]
  \centering
  \includegraphics[width=\textwidth]{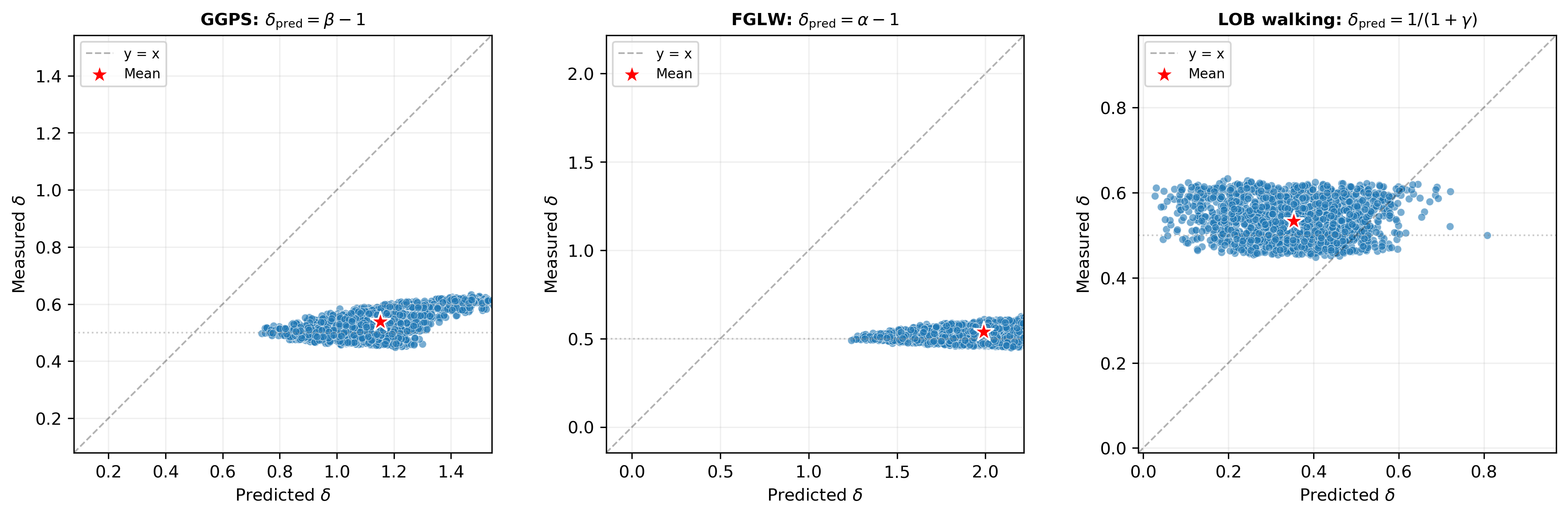}
  \caption{Predicted vs.\ measured $\delta$ for the same 2000 baseline
    stocks (square panels, equal axes).
    Horizontal axis: measured $\delta$; vertical axis: predicted $\delta$.
    Left: GGPS ($\delta_{\text{pred}} = \beta - 1$); centre: FGLW
    ($\delta_{\text{pred}} = \alpha - 1$); right: LOB walking
    ($\delta_{\text{pred}} = 1/(1+\gamma)$).
    Diagonal: perfect agreement. Red star: cross-stock mean.
    All points in the left and centre panels lie above the diagonal
    (over-prediction); in the right panel most lie below
    (under-prediction).}
  \label{fig:theory}
\end{figure}

Since each of the three quantitative predictions can be evaluated on the
same per-stock data, Figure~\ref{fig:theory} offers a direct,
model-internal test: GGPS and FGLW predict $\delta > 1$ in every stock
(left and centre panels, all points above the diagonal), while LOB
walking predicts $\delta < 0.5$ in most stocks (right panel, below the
diagonal). None of the three passes through the cloud of measured
$\delta\approx 0.5$. The failure of all three single-mechanism
predictions motivates the counterfactual experiments of
Section~\ref{sec:counterfactual}, which identify the mechanisms that
\emph{are} responsible.

\subsection{Counterfactual experiments}
\label{sec:counterfactual}

Table~\ref{tab:counterfactual} collects the results of eight
counterfactual scenarios, each a variant of the baseline in which one
mechanism is suppressed or altered. All scenarios run 20 independent stocks
(see Section~\ref{sec:model}).

\begin{table}[H]
\centering
\caption{Counterfactual results (20 independent stocks, cross-stock mean $\pm$ SE).
$\delta$: fitted impact exponent; SE = std$/\sqrt{N_{\text{stocks}}}$.
$\delta_{\text{GGPS}} = \beta - 1$,
$\delta_{\text{FGLW}} = \alpha - 1$,
$\delta_{\text{LOB}} = 1/(1+\gamma)$.
Only institutional agents (agent\_id $< N_{\text{inst}} = 20$) included.}
\label{tab:counterfactual}
\begin{tabular}{lrrrrrrr}
\toprule
Scenario & $\delta$ & $c$ & $N_{\text{meta}}$ ($\times 10^3$) &
  $\delta_{\text{GGPS}}$ & $\delta_{\text{FGLW}}$ & $\delta_{\text{LOB}}$ & $\gamma$ \\
\midrule
Baseline       & $0.549 \pm 0.013$ & 0.98 & 380 & 1.23 & 2.11 & 0.29 & 2.46 \\
No splitting   & $0.324 \pm 0.015$ & 0.22 & 236 & 1.12 & 1.49 & 0.38 & 1.64 \\
No HFT         & $0.386 \pm 0.008$ & 0.51 & 345 & 1.05 & 3.65 & 0.20 & 4.08 \\
Price limits   & $0.529 \pm 0.014$ & 0.87 & 271 & 1.24 & 2.12 & 0.38 & 1.60 \\
Low liquidity  & $0.543 \pm 0.012$ & 1.04 & 1251 & 1.19 & 2.09 & 0.42 & 1.39 \\
Momentum       & $0.549 \pm 0.013$ & 0.98 & 342 & 1.23 & 2.10 & 0.30 & 2.37 \\
Uniform split  & $0.547 \pm 0.014$ & 0.97 & 381 & 1.22 & 2.09 & 0.26 & 2.82 \\
Front-loaded   & $0.502 \pm 0.014$ & 0.69 & 372 & 1.19 & 2.10 & 0.26 & 2.82 \\
\midrule
Empirical (TSE) & 0.489 & 0.842 & --- & --- & --- & --- & $\sim$1.0 \\
\bottomrule
\end{tabular}
\end{table}

The table shows a clear hierarchy. Removing order splitting causes
$\delta$ to collapse from $0.549$ to $0.324$, the largest single change
among all scenarios. Without splitting, every metaorder lands as one huge
market order that consumes the available depth at a single instant; the
LOB has no time to recover between slices. Removing HFT agents has a
similar direction ($\delta \to 0.386$) but a weaker effect: the book still
receives some resting liquidity from retail limit orders, but without
active replenishment the depth is thinner and impact is larger.

\begin{figure}[H]
  \centering
  \includegraphics[width=\textwidth]{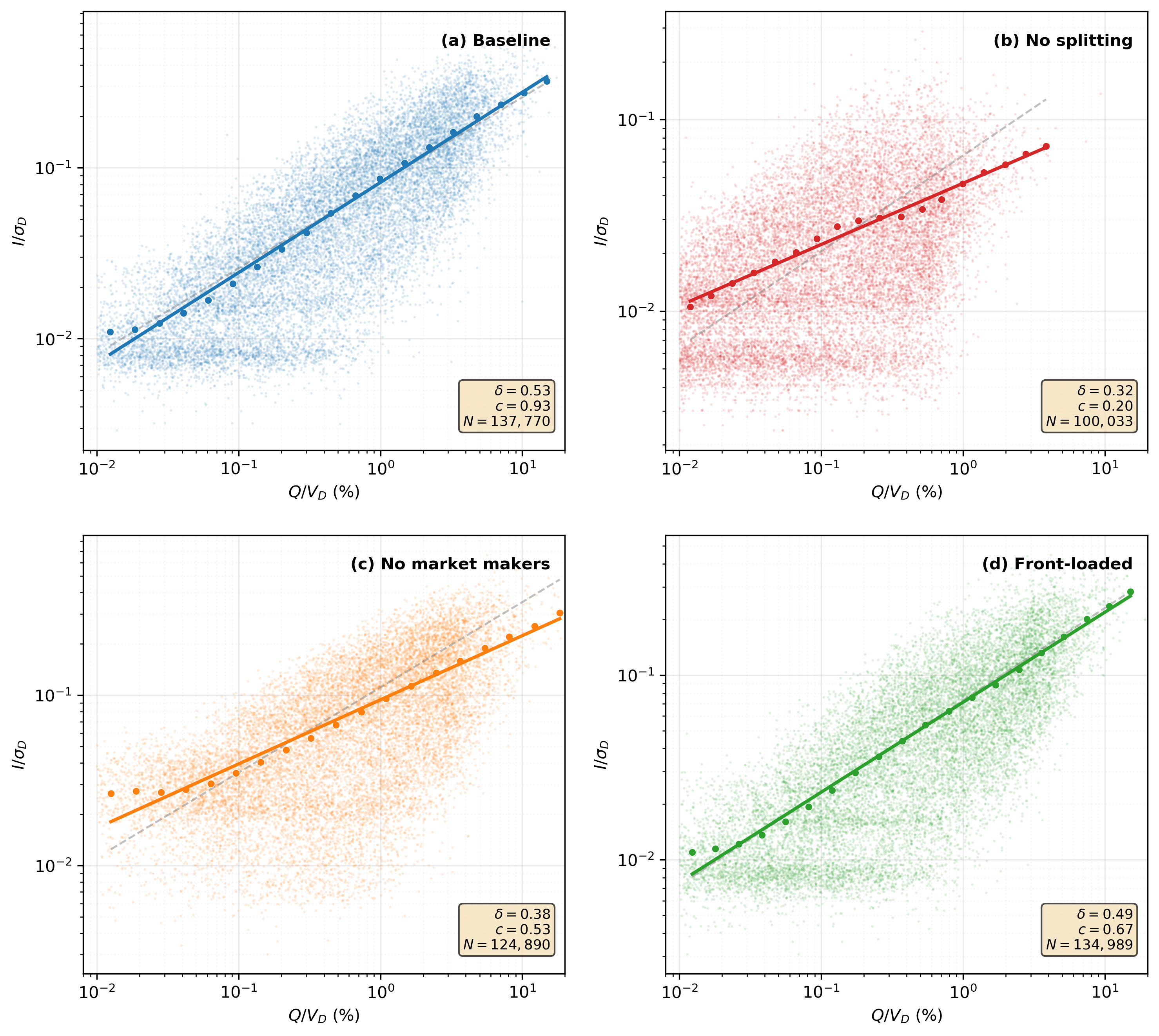}
  \caption{Impact curves (log-log) for four selected scenarios.
    Same conventions as Figure~\ref{fig:per_stock}.
    Baseline~(a): clear concave power law ($\delta = 0.55$).
    No splitting~(b): the impact curve flattens ($\delta = 0.32$); note
    the absence of a pure power law at large $Q/V_D$.
    No HFT~(c): steeper than baseline but still concave ($\delta = 0.39$).
    Front-loaded~(d): nearly indistinguishable from~(a) ($\delta = 0.50$).}
  \label{fig:counterfactual_curves}
\end{figure}

\begin{figure}[H]
  \centering
  \includegraphics[width=\textwidth]{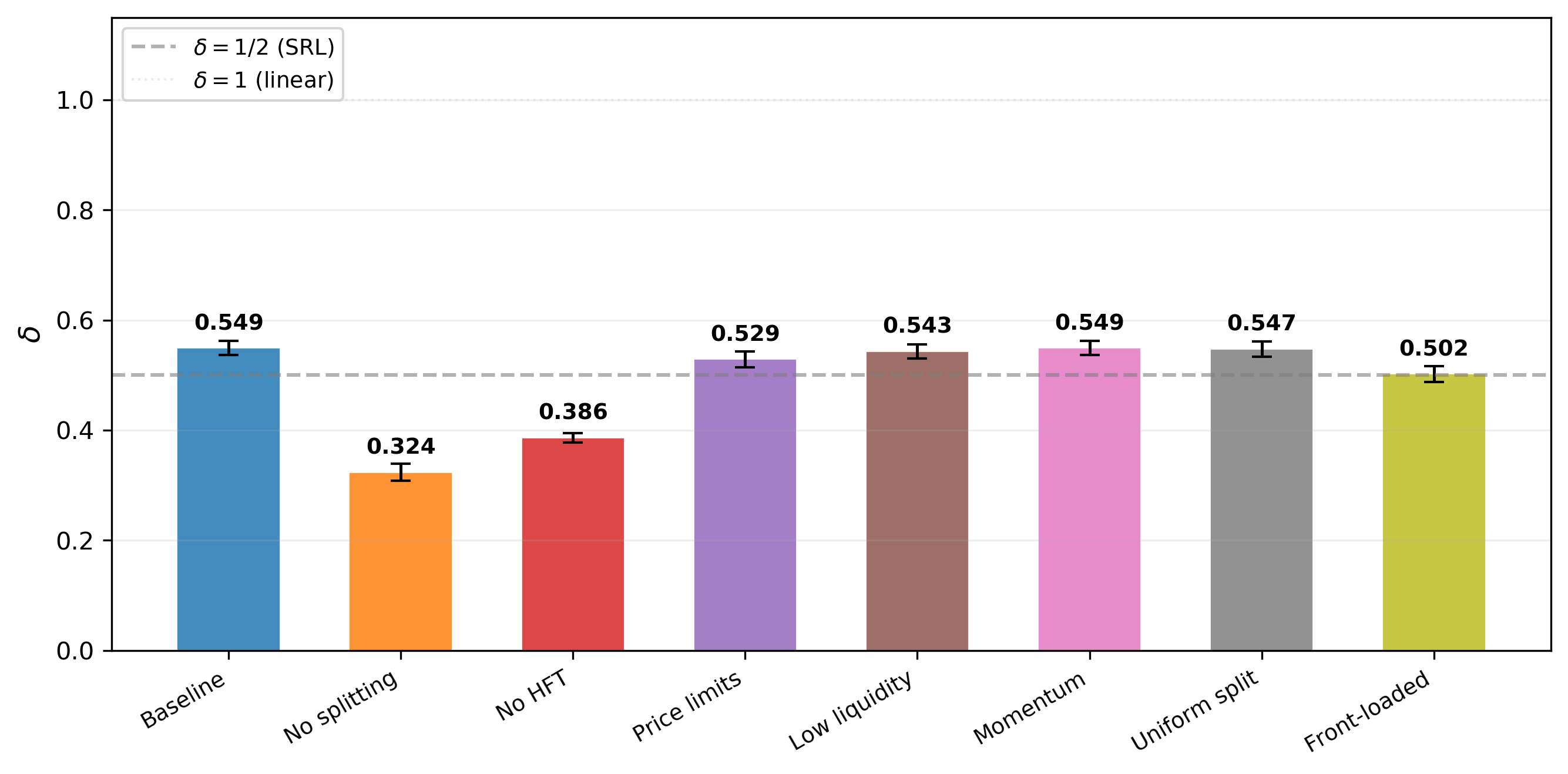}
  \caption{Fitted $\delta$ for each counterfactual scenario.
    Dashed horizontal: $\delta = 1/2$; dotted: $\delta = 1$ (linear).
    Error bars: cross-stock SE $= \text{std}/\sqrt{N_{\text{stocks}}}$.}
  \label{fig:counterfactual}
\end{figure}

By contrast, perturbations that leave splitting and HFT intact barely
move $\delta$. Price limits, momentum trading, reduced background
liquidity, uniform splitting, and front-loaded splitting all yield
$\delta\in[0.49, 0.53]$ --- within 10\% of the baseline.
Figure~\ref{fig:counterfactual_curves} shows the impact curves for the
four most illustrative scenarios, and Figure~\ref{fig:counterfactual}
summarises the fitted $\delta$ across all eight scenarios.
The baseline curve is concave
($\delta = 0.55$); the no-splitting curve is nearly flat
($\delta = 0.32$).

\subsection{Theory stress test}

We now examine in detail why each of the three predictions fails on the
baseline data (Figure~\ref{fig:theory}).

GGPS ($\delta=\beta-1$) over-predicts in every stock
($\delta_{\text{GGPS}}\in[0.74, 1.56]$, mean $1.16$), because the Hill
estimator $\beta$ captures the unconditional size tail whereas impact
depends on the \emph{conditional} displacement given size. FGLW
($\delta=\alpha-1$) fares even worse ($\delta_{\text{FGLW}}\in[1.26,
2.47]$, mean $2.00$); both theories predict $\delta > 1$, far above the
observed $\delta\approx 0.5$. LOB walking ($\delta=1/(1+\gamma)$) goes in
the opposite direction: the fitted $\gamma\in[0.5, 13]$ is much larger
than the empirical $\gamma\approx 1$, so the prediction
$\delta_{\text{LOB}}\in[0.07, 0.69]$ (mean $0.36$) systematically
underestimates the measured $\delta$. The visible book in our model is too
concave, and the latent liquidity that Donier et al.~\citep{donierFullyConsistentMinimal2015}
argue softens impact is not captured by a static snapshot of the LOB.

No single theory reproduces $\delta$ within our model (Figure~\ref{fig:theory}).
The SRL appears to arise from the coupling of splitting (which creates
time gaps) and liquidity replenishment (which refills the book during
those gaps) --- the picture suggested by the latent liquidity framework.
We note that this is a computational, model-internal test; whether the
same conclusion holds in real markets would require analogous empirical
measurement.

%% =====================================================================
\section{Discussion}
%% =====================================================================

\subsection{Necessary mechanisms for the SRL}

Two mechanisms stand out as necessary. Order splitting is the dominant
factor: its removal collapses $\delta$ from $0.549$ to $0.324$, well below
both $1/2$ and $1$. The microscopic order flow explains this. In the
baseline, a metaorder of size $Q=10^4$ is split
into, say, $N_c=50$ child orders arriving every $\lambda=3$ ticks. Between
children, HFT agents replenish the book at the best quotes. The cumulative
impact is the sum of $N_c$ small kicks, each partially absorbed by fresh
liquidity. When splitting is turned off, the same $Q=10^4$ hits the book
as a single market order. The LOB is traversed from the best bid (or ask)
outward, and the total price displacement is set by the cumulative depth
profile $V(q)$ at the single instant of arrival. Because $V(q)$ grows
super-linearly in our model ($\gamma\approx 2$), larger unsplit orders
encounter disproportionately more depth per unit of additional size, and
the resulting impact curve is concave with $\delta < 1/2$. The precise
value $\delta = 0.324$ reflects the specific shape of the resting book
at the moment of arrival --- it is not a universal constant but a
consequence of the model's LOB geometry.

Liquidity replenishment by HFT agents is the second necessary ingredient.
When HFT is removed, the book still receives resting orders from retail
limit submissions, but the active two-sided quoting that refills depleted
levels after each fill is gone. The LOB thins during execution, so each
subsequent child order moves the price further than it would in a
replenished book, producing a steeper impact curve ($\delta = 0.386$).

Changing the splitting rule (Dirichlet, uniform, or front-loaded) has
virtually no measurable effect ($\delta\in[0.49, 0.53]$). Adding momentum traders,
imposing price limits, or halving the retail population likewise leave
$\delta$ within $10\%$ of the baseline. This robustness is consistent
with the empirical observation that $\delta\approx 0.5$ holds across
markets with widely varying trading rules and participant
mixes~\citep{bouchaudFluctuationsResponseFinancial2004, moroMarketImpactTrading2009, tothAnomalousPriceImpact2011}.

\subsection{Relation to theoretical frameworks}

\begin{figure}[H]
  \centering
  \includegraphics[width=\textwidth]{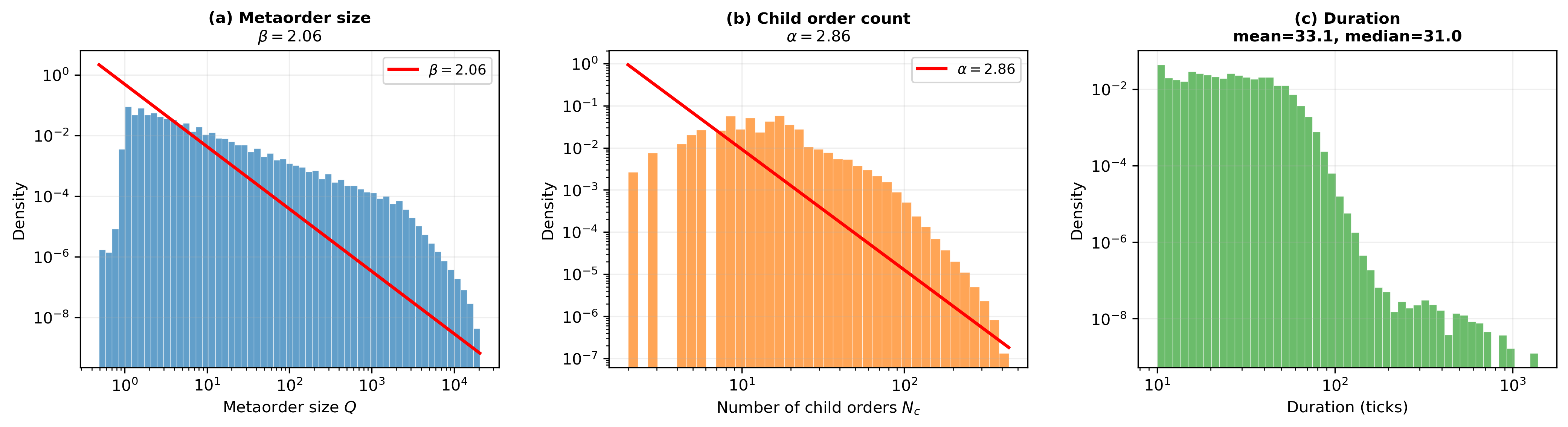}
  \caption{Baseline metaorder statistics (single representative stock, institutional
    agents only).
    (a)~Complementary CDF of metaorder size $Q$ (log-log); the tail
    follows $Q^{-\beta}$ with $\beta = 2.10$ (dashed line).
    (b)~Complementary CDF of child-order count $N_c$; tail exponent
    $\alpha = 2.81$.
    (c)~Duration histogram (log-log scale); mean $= 132$ ticks,
    median $= 78$ ticks. The right tail extends beyond $10^3$ ticks.}
  \label{fig:metaorder_stats}
\end{figure}

Three analytical frameworks predict $\delta$ from exponents that can be
measured independently. We sketch each derivation's logic, identify its
core assumption, and explain why that assumption fails in our model.

\paragraph{GGPS: $\delta = \beta - 1$.}
Gabaix et al.~\citep{gabaixInstitutionalInvestorsStock2006} argue as follows.
If metaorder sizes follow a power law $P(Q > q) \sim q^{-\beta}$ and
each unit of volume moves the price by a constant amount $\lambda$,
then a metaorder of size $Q$ causes impact $I = \lambda Q$.
The largest metaorder on a given day has $Q_{\max} \sim V_D^{1/\beta}$
(because $P(Q > Q_{\max}) \sim 1$), so the daily price range
$\sigma_D \sim \lambda V_D^{1/\beta}$.
Eliminating $\lambda$ from $I/Q = \lambda = \sigma_D / V_D^{1/\beta}$
and normalising by $\sigma_D$ gives
\begin{equation}
  \frac{I}{\sigma_D} = \frac{Q}{V_D^{1/\beta}\,\sigma_D}
  \cdot \sigma_D = \left(\frac{Q}{V_D}\right)^{1}
  V_D^{1-1/\beta}\,.
\end{equation}
Matching this against the empirical form
$I/\sigma_D = c\,(Q/V_D)^{\delta}$ requires
$\delta = 1$ and $c \sim V_D^{1/\beta - 1}$ ---
but this only reproduces the linear law. To obtain
$\delta < 1$, GGPS posit that large trades find progressively
more hidden liquidity, effectively diluting per-unit impact to
$\lambda \sim Q^{\beta - 2}$. Substituting into the normalisation
yields $\delta = \beta - 1$.

\emph{Why it fails.}
The prediction requires a pure power-law tail $P(Q>q)\sim q^{-\beta}$
from which $\beta$ can be estimated. In our simulation the metaorder
size distribution (Figure~\ref{fig:metaorder_stats}, panel~a) is only
approximately Pareto --- the tail bends due to the bounded
$[5,50]$ child-order count --- so the Hill estimator gives a
well-defined $\beta$ but its physical meaning as a tail exponent is
questionable. Even granting the exponent, the hidden-liquidity dilution
$\lambda \sim Q^{\beta-2}$ is an ad hoc assumption: it posits that
per-unit impact decreases with order size, but provides no mechanism.
In our model the dilution arises not from a size-dependent $\lambda$
but from order splitting --- a metaorder of size $Q$ is broken into
$N_c$ small slices, each encountering a quasi-stable book. The
splitting mechanism, not the tail exponent, determines
$\delta$ --- consistent with Zarinelli et al.'s finding that
impact depends on participation rate and execution duration,
not size alone~\citep{zarinelliSquareRootEvidence2015}.

\paragraph{FGLW: $\delta = \alpha - 1$.}
Farmer et al.~\citep{farmerHowEfficiencyShapes2013} replace the size
distribution with the distribution of child-order counts $N_c$.
If $P(N_c > n) \sim n^{-\alpha}$ and each child order displaces the
mid-price by $\Delta p \sim \sigma_D / V_D$ (one ``tick'' of
impact), then the total impact $I \sim N_c\,\Delta p$. The same
normalisation argument as GGPS then gives $\delta = \alpha - 1$.

\begin{figure}[H]
  \centering
  \includegraphics[width=\textwidth]{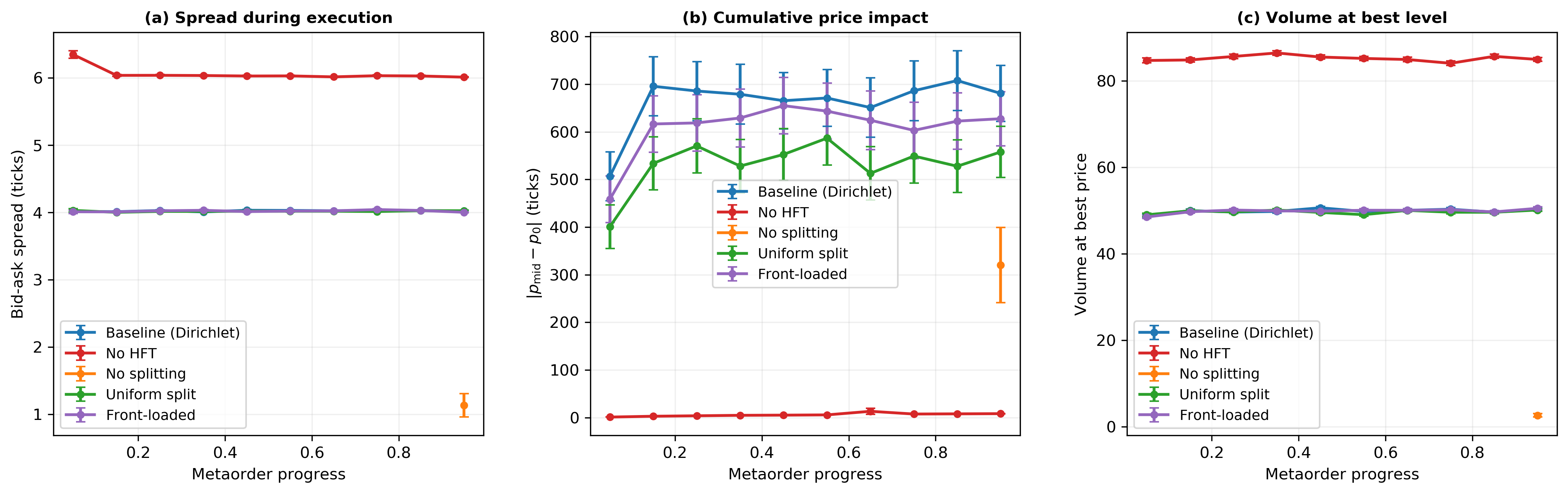}
  \caption{LOB dynamics during metaorder execution across five scenarios.
    (a)~Bid-ask spread vs.\ metaorder progress.
    (b)~Cumulative price impact $|p_{\mathrm{mid}} - p_0|$ vs.\ progress.
    (c)~Volume at the best price level vs.\ progress.
    Error bars: $\pm 1$ SE across metaorders.}
  \label{fig:lob_evolution}
\end{figure}

\emph{Why it fails.}
As with GGPS, the prediction relies on a well-defined power-law tail for
$N_c$ (Figure~\ref{fig:metaorder_stats}, panel~b), but the bounded
range $[5,50]$ imposed by the splitting algorithm produces a
distribution that deviates from a pure power law. Beyond the
distributional issue, the assumption $\Delta p \sim \text{const}$
means every child order moves the price by the same amount. This would
be correct if the LOB were static and all children had identical size.
But the child volumes are drawn from a Dirichlet distribution (so they
vary), and although HFT keeps the book quasi-stable
(Figure~\ref{fig:lob_evolution}), the mid-price drifts as execution
proceeds, so later children encounter a shifted book. This breaks the
proportionality $I \propto N_c$.

\paragraph{LOB walking: $\delta = 1/(1+\gamma)$.}
If the visible cumulative depth grows as $V(q) \sim q^{1+\gamma}$
with distance $q$ from the mid-price, a single market order of size
$Q$ must walk through the book until $V(q^\star) = Q$. Solving
$q^{1+\gamma} = Q$ gives $q^\star \sim Q^{1/(1+\gamma)}$, so the
price displacement is
\begin{equation}
  I \sim q^\star \sim Q^{1/(1+\gamma)},
  \qquad \delta = \frac{1}{1+\gamma}\,.
\end{equation}

\emph{Why it fails.}
The formula describes a single market order of size $Q$ crossing the
book at one instant. In our model, by contrast, a metaorder is split
into $N_c$ child orders spaced by Poisson gaps. HFT agents replenish
depleted levels between children (Figure~\ref{fig:lob_evolution}), so
no single child order walks far into the book. The cumulative impact
is the sum of many small displacements, each absorbed by fresh
liquidity, rather than the single traversal assumed by the formula.
Our model produces $\gamma\in[1.5,3.6]$ (concentrated near-best-quote
liquidity), predicting $\delta\in[0.2,0.4]$ for a single-order walk.
The measured $\delta\approx 0.5$ is higher because splitting prevents
the book from being walked through in one pass.

\paragraph{Summary.}
The three theories fail for two reasons. First, the distributional
assumptions do not hold: the simulated metaorder size and child-order
count distributions are only approximately power-law
(Figure~\ref{fig:metaorder_stats}), so the exponents $\beta$ and
$\alpha$ are not well-defined tail indices. Second, all three theories
describe the impact of a single order of size $Q$ on a given book
state --- they lack a mechanism for \emph{split execution}. In our
model, order splitting breaks a large trade into many small child
orders, and HFT replenishment keeps the book quasi-stable between them
(Figure~\ref{fig:lob_evolution}). The cumulative impact is the sum of
many small displacements absorbed by fresh liquidity, not a single
traversal of the book. Within our model, this coupling of splitting and replenishment is the only
mechanism that survives counterfactual ablation. It is consistent with the
latent-liquidity picture of Donier et al.~\citep{donierFullyConsistentMinimal2015},
in which a hidden liquidity profile continuously revealed by price
diffusion yields $\delta = 1/2$ independently of the visible $\gamma$;
what we add is the agent-level identification of which components are
causally necessary, obtained by removing each one in turn.

\subsection{Relation to earlier ABM work}

Several agent-based frameworks have reproduced the square-root law.
T\'{o}th et al.~\citep{tothAnomalousPriceImpact2011}
showed that $\delta \approx 1/2$ follows from the interplay between
sequential metaorder execution and the shape of the latent order book.
Their analytical framework derives a locally linear supply/demand
profile from the assumption of diffusive prices and Poisson order flow,
confirmed by a numerical model.
However, the framework does not decompose the mechanism into
agent-level ingredients: there are no designated market makers or
explicit splitting strategies, and the model cannot identify which
components are causally necessary for $\delta \approx 1/2$.
Mastromatteo et al.~\citep{mastromatteoAgentbasedModelsLatent2014} extended
the $\varepsilon$-intelligence framework with stimulated liquidity refill
--- where limit-order arrival responds to recent market-order activity ---
and confirmed concave impact across execution protocols, but did not
decompose the mechanism into separable agent-level ingredients.
Donier et al.~\citep{donierFullyConsistentMinimal2015} derived $\delta = 1/2$
analytically from a latent liquidity framework, providing a
continuum theory that bypasses agent-level detail entirely.

Our model differs from these in three respects.
First, it retains explicit institutional structure --- metaorders are
generated by dedicated agents that split large parent orders into
sequences of child orders arriving as a Poisson process, while HFT
agents provide two-sided liquidity with adaptive replenishment.
This makes it possible to ablate each component independently
(remove splitting, remove HFT, etc.) and measure the causal effect on
$\delta$ --- something a zero-intelligence model~\citep{farmerPredictivePowerZero2005} with homogeneous agents cannot do.
Second, the four agent types (institutional, HFT, retail, news) are
motivated by real market ecology rather than mathematical convenience.
The resulting order flow exhibits stylised facts --- heavy-tailed
returns, clustered volatility, concave impact --- without being
calibrated to any particular dataset.
Third, we benchmark three competing quantitative predictions
(GGPS, FGLW, LOB walking) on the same simulated data, treating the
model as a computational stress test that complements the empirical
approach of~\citep{bouchaudHowMarketsSlowly2008,
zarinelliSquareRootEvidence2015, bucciCoimpactCrowdingEffects2019}.
The contribution is therefore methodological rather than model-based:
the ABM is a tool for causal dissection, not a proposed market model
in its own right.

\subsection{Limitations}

The tick-level return ACF(1) $\approx -0.35$ is a genuine bid--ask
bounce signature~\citep{rollSimpleImplicitMeasure1984}, but the model also exhibits weak
longer-horizon mean reversion because it lacks an exogenous fundamental
price. The visible LOB shape exponent $\gamma\in[1.5, 3.6]$ is higher
than the empirical $\gamma\approx 1$, a consequence of the minimal agent
ecology (see discussion above). Adding a richer mix of order-placement
strategies could bring $\gamma$ closer to 1, but doing so would also
diminish the discriminatory power of the LOB walking stress test. Each stock is simulated in isolation, ignoring
cross-asset correlations. Our metaorders are reconstructed ex post from
consecutive same-agent trades, whereas empirical studies use proprietary
execution data; the two identification procedures need not yield the
same size distributions.

%% =====================================================================
\section{Conclusion}
%% =====================================================================

A minimal limit-order-book ABM with four heterogeneous agent types
reproduces the square-root law of market impact
($\langle\delta\rangle = 0.539$ across 2000 stocks, vs.\
$\langle\delta\rangle = 0.489$ on the TSE). Two lines of evidence
identify the responsible mechanisms. First, counterfactual ablation shows
that only the removal of order splitting ($\delta\to 0.324$) or of
liquidity replenishment by market makers ($\delta\to 0.386$) breaks the
SRL, while perturbations leaving both intact move $\delta$ by less than
$10\%$. Order splitting and liquidity replenishment are thus jointly
necessary for the SRL within this model; the splitting rule, momentum
trading, price limits, and background liquidity level are dispensable.
Second, a per-stock comparison of three theoretical predictions (GGPS,
FGLW, LOB walking) against simulated $\delta$ rejects all three on
identical data: distribution-based theories over-predict,
visible-book-based theories under-predict. The failure of all three
single-mechanism predictions, together with the ablation results,
supports the view that the SRL arises from the coupling of splitting
and replenishment, consistent with the latent liquidity
picture~\citep{mastromatteoAgentbasedModelsLatent2014, donierFullyConsistentMinimal2015}.

%% =====================================================================
\section*{Data availability}
%% =====================================================================

The simulation code and analysis scripts are available from the
corresponding author on reasonable request. All figures in this paper
were generated from agent-based simulations run on a 40-core server;
the raw trade records and fitted parameters can be shared for
reproduction purposes.

The model uses four agent types with the following default parameters:
$N_{\text{inst}}=20$ institutional agents (Pareto shape $\xi\sim\text{Uniform}[1.5,3.5]$,
child-count range $[5,50]$, Poisson gap rate $\lambda=3$),
$N_{\text{HFT}}\in[3,10]$ high-frequency traders,
$N_{\text{retail}}\in[100,400]$ retail agents, and $N_{\text{news}}=3$ news agents.
Each stock runs $10^6$ time steps (200 trading days,
5000 steps/day) with tick size $=1$ and initial mid-price $=10\,000$.
The model is minimal-intelligence: agents follow fixed behavioural rules
with no learning, optimisation, or strategic adaptation.

%% =====================================================================
\section*{Acknowledgements}
%% =====================================================================

This work was supported by the Scientific Research Project
(No.WU2025B011) and the Start-up Funding of Westlake University.

%% =====================================================================
\section*{Declaration of generative AI and AI-assisted technologies
in the manuscript preparation process}
%% =====================================================================

During the preparation of this work the author used Claude
(Anthropic) in order to improve language readability and check grammar.
After using this tool, the author reviewed and edited the content as
needed and takes full responsibility for the content of the published
article.

%% =====================================================================
%% References
%% =====================================================================

\bibliographystyle{elsarticle-num}
\bibliography{references}

\end{document}